\documentclass[12pt]{iopart}

\usepackage{setspace}
\usepackage{latexsym}
\usepackage{amssymb}
\usepackage{graphicx,color}
\usepackage{epstopdf}
\usepackage{graphicx}
\usepackage{setspace}
\usepackage{makeidx}

\def\nurho{\nabla^{\rho}}
\def\ndrho{\nabla_{\rho}}
\def\nukappa{\nabla^{\kappa}}
\def\ndkappa{\nabla_{\kappa}}

\def\nualpha{\nabla^{\alpha}}
\def\ndalpha{\nabla_{\alpha}}
\def\nubeta{\nabla^{\beta}}

\def\ndlambda{\nabla_{\lambda}}

\def\tndalpha{\widetilde{\nabla}_{\alpha}}

\def\tnukappa{\widetilde{\nabla}^{\kappa}}

\def\pdkappa{\partial_{\kappa}}
\def\pukappa{\partial^{\kappa}}

\def\no{\nonumber}

\def\de{\delta}

\def\imj{\int_{M}d^{n}x\,\sqrt{-g}}
\def\isj{\int_{\Sigma}d^{n-1}x\,\sqrt{-h}}
\def\ime{\int_{M}d^{n}x\,\sqrt{-\tilde{g}}}
\def\ise{\int_{\Sigma}d^{n-1}x\sqrt{-\tilde{h}}}

\def\bea{\begin{eqnarray}}  
\def\eea{\end{eqnarray}}

\def\be{\begin{equation}}
\def\ee{\end{equation}}

\makeatletter

\newcommand{\Rmnum}[1]{\expandafter\@slowromancap\romannumeral #1@}
\makeatother

\begin{document}

\title{The dynamical equivalence of modified gravity revisited}
\author{Ippocratis D. Saltas and Mark Hindmarsh} 
\address{Department of Physics \& Astronomy, University of Sussex, Brighton, BN1 9QH, United Kingdom} 
\ead{ I.Saltas@sussex.ac.uk {\rm and} M.B.Hindmarsh@sussex.ac.uk} 
\begin{abstract}
We revisit the dynamical equivalence between different representations of vacuum modified gravity models in view of Legendre transformations. The equivalence is discussed for both bulk and boundary space, by including in our analysis the relevant Gibbons--Hawking terms. In the $f(R)$ case, the Legendre transformed action coincides with the usual Einstein frame one. We then re-express the $R+f(G)$ action, where $G$ is the Gauss--Bonnet term, as a second order theory with a new set of field variables, four tensor fields and one scalar and study its dynamics. For completeness, we also calculate the conformal transformation of the full Jordan frame $R+f(G)$ action. All the appropriate Gibbons--Hawking terms are calculated explicitly. 
\end{abstract}

\pacs{04.20.Cv, 04.20.Fy, 04.50.Kd}
\submitto{Class. Quantum Grav.}
\maketitle
\section{Introduction}
From the early development of modified (higher order) gravity models \cite{StarobinskyInflation1}, their dynamical equivalence to different representations has been an issue of research and intense debate. What the term ``dynamical equivalence" means is that a particular gravitational action can be re-expressed as a new one, with a new set of field variables, and that there is an invertible mapping that relates the two sets of field variables, as well as the two actions (or Lagrangians) respectively. The variational principle of the new action will in principle require different boundary conditions and possibly different Gibbons--Hawking terms as well. Probably the most well known example of such an equivalence is that of the $f(R)$ action to Brans--Dicke and Einstein--Hilbert one, the first through the introduction of an auxiliary scalar field \cite{Higgs_Auxiliary_Scalars}--\cite{Wands_Auxiliary_Scalars}, and the second through a conformal transformation of the metric \cite{Higgs_Auxiliary_Scalars, Teyssandier&Tourrenc_Auxiliary_Scalars, Whitt_Auxiliary_Scalars}, \cite{MagnanoEtal_CT_Einstein_Frame}--\cite{Cotsakis&Barrow_CT_Einstein_Frame}. 

The motivation of re-expressing a gravitational action by introducing a new set of field variables might be related mainly to two things. The first is mathematical simplicity and convenience, if the new set of variables is to make the calculations one wants to perform simpler. Furthermore, if one is able to move from a variational principle that will lead to fourth order equations of motion to another that will lead to second order ones, that could be a benefit, since second order equations are in principle easier to handle as well as to interpret physically. The second possible motivation is related to physics. One gets more intuition and understanding of a gravitational theory, by studying its equivalence to other ones (like the equivalence between $f(R)$ and Einstein gravity).

From a physical point of view care must be taken in the interpretation of physical quantities in the two different representations. The question that often arises is which of the two representations is the physical one. For example, in the case of the conformal equivalence between $f(R)$ and Einstein gravity, the inclusion of matter in the action can raise the question of along which of the two metrics (original and conformally transformed one) do matter particles actually fall. For some interesting discussions on the subject one can refer to Refs \cite{Brans_Metric_Significance}--\cite{SotiriouEtal_No_Progress}. 

Given the equivalence between the bulk parts of two actions, this does not mean that the equivalence holds for the surface parts as well. More precisely, given the GH term of an action in one representation, then the GH term calculated using the equivalence with the other reresentation is possible to be problematic as far as the particular variational principle is concerned. As we will discuss later on, this is the case for $f(R)$ and $R+f(G)$ gravity, when the equivalence between the original and the Jordan frame action is considered. As we will see, the latter equivalence breaks on the boundary surface.

In this paper, we will focus on $f(R)$ \cite{Nojiri&Odintsov_Review, Capozziello&Francaviglia_f(R)_Review, f(R)SotiriouFaraoni, f(R)Felice&Tsujikawa} and $R + f(G)$ \cite{Nojiri&Odintsov_Review, Nojiri&Odintsov_R+f(G), Nojiri&Odintsov_GB_DE} models, where $G$ is the Gauss--Bonnet (GB) term defined as \cite{Lovelock}
\begin{equation}
G \equiv R^2-4R^{\alpha \beta}R_{\alpha \beta}+R^{\alpha}{}_{\beta \gamma \delta}R_{\alpha}{}^{\beta \gamma \delta}, \label{GB}
\end{equation}
$``f"$ being an in principle non linear function of its arguments. We will study the dynamical equivalence of above theories to other representations in vacuum, using as our main tool the rather general approach of Legendre transformation, which for the case of $f(R)$ coincides with the standard procedure of introducing an auxiliary scalar field followed by a conformal transformation, something which is not true for $R + f(G)$ theories, i.e. the latter cannot be conformally transformed to a minimally coupled, scalar-tensor frame. However, for the $R + f(G)$ theory, we will show how using a Legendre transformation we can re-express it as a second order theory, with a new extra rank two tensor field. We will work at the level of the action and we will include in our analysis the relevant Gibbons--Hawking terms \cite{Gibbons&Hawking}, which are important for the consistency of the initial value formulation of the theory. Furthermore, we will calculate them explicitly wherever necessary. 

The structure of the paper is as follows: In Section  \ref{LagrangianFormulation} we briefly describe the approach of Legendre transforming a higher order gravitational action. In view of the latter approach, in Section \ref{f(R)1} we show the equivalence of the full (including the relevant GH term) $f(R)$ action to the Einstein--Hilbert one. We also discuss the equivalence between the relevant GH terms. Then, in Section \ref{f(G)1} we consider the $R + f(G)$ action and after calculating the GH term in the Jordan (scalar--tensor) frame, we explicitly study the effect of conformal transformation on the full Jordan frame action. Finally, in Section \ref{f(G)2} we re-express the original $R + f(G)$ action with one scalar and two extra tensor fields present (apart from the metric), one of them though being independent, and then discuss the classical dynamics of the system. Useful formulas and calculations can be found in the Appendix. We will work in natural units, $c = G = 1$.

\section{Establishing dynamical equivalence} \label{LagrangianFormulation}
In this section we will briefly describe the idea and motivation behind Legendre transforming a gravitational action as a tool of moving to a new, dynamically equivalent variational principle of second-order, first applied in Refs \cite{Magnano1, Magnano2}. 

The higher order nature of non--linear gravitational Lagrangians like the $f(R)$ or $R + f(G)$ ones, comes from the fact that they are non--linear with respect to the second derivatives of the metric, since ``$f$" is an in principle non--linear function with respect to its arguments, and $R^{\alpha}{}_{\beta \gamma \delta} = R^{\alpha}{}_{\beta \gamma \delta} \left( g^2,(\nabla g)^2, \nabla^2 g \right)$. 

However, we can try to re--express higher order gravitational Lagrangians linearly with respect to $\nabla^2 g$, by making them linear with respect to the curvature tensors \footnote{This is because curvature tensors are linear with respect to $\nabla^{2}g$.} through the introduction of the appropriate ``velocities" and ``momenta", in a similar fashion to the ordinary Hamiltonian formalism. The new variational principle will then lead to second order equations of motion for the new set of field variables. 

The appropriate identification for the generalised ``position" and ``moment" as well as the Legendre transformed Lagrangian will read schematically as 
\begin{eqnarray}
q \leftrightarrow g_{\alpha \beta}, \; \; \;  \dot{q_{i}} \leftrightarrow \left( R, R^{\alpha \beta}, R^{\alpha}{}_{\beta \gamma \delta} \right), \label{GeneralisedVelocitiesGeneral}
\end{eqnarray}
\be
\widetilde{L} {(-g)}^{-1/2} =  \dot{q}_{i}p_{i} -  \left[  \dot{q}_{i}(q,p)p_{i} - {(-g)}^{-1/2}L(q,p)  \right] \equiv \dot{q}_{i}p_{i} - H(q,p),
\ee
assuming the invertibility condition holds, $\partial^2 L/( \partial \dot{q}_{i}\partial \dot{q}_{j} ) \neq 0$. Quantities entering $\widetilde{L} $ will be in principle tensor objects. Different gravity actions give us different options in defining generalised ``velocities" and ``momenta". This will be made clear in Sections \ref{f(R)2} and \ref{f(G)2}, where we will apply the above formalism for the case of $f(R)$ and $R+f(G)$ gravity respectively. 

\section{Dynamical equivalence of $f(R)$ gravity, part \Rmnum{1}} \label{f(R)1}
It is well known in the literature that through the introduction of an auxiliary scalar, the $f(R)$ action can be re-expressed as a non--minimally coupled scalar-tensor one (also called Jordan frame action), and it is the latter that is usually conformally transformed to the so-called Einstein frame action.
In this section, we we will focus and attempt to clarify the role of the relevant GH terms in the two representations, original and Jordan frame one. Then, in the next subsection we will demonstrate the equivalence of the full action (bulk and surface part) using the general approach of the Legendre transformation.
Starting from the bulk $f(R)$ action on a manifold $M$,
\be
S = \imj f(R) , \label{f(R)_action}
\ee
through the introduction of an auxiliary scalar $\psi$, one can re-express it in a dynamically equivalent way as
\be
S^{\rm{J}} =  \imj \left[ \Phi R(g)-V(\Phi)  \right],  \label{f(R)HJ}
\ee
with $ \Phi \equiv f'(\psi)$, and $V(\Phi) \equiv  \Phi f'^{-1}(\Phi)-f\left( f'^{-1}(\Phi) \right)$. For the latter we require that $f''(\psi) \neq 0$, so we are able to solve for $\psi = f'^{-1}(\Phi)$. Action (\ref{f(R)HJ}) is the so-called Jordan frame action.

The transition to the Einstein frame action will be shown explicitly in Section \ref{f(R)2} by means of a Legendre transformation.
\subsection{The $f(R)$ Gibbons--Hawking term in the Jordan frame} \label{f(R)1.1}
When considering gravitational actions on manifolds with boundary $\Sigma$, the variation gives boundary terms containing normal derivatives of the metric variation $\nabla_{n}^{(k)} \delta g_{\alpha\beta}$. However, a well defined variational principle requires that only a particular set of dynamical coordinates ($g_{\alpha\beta}$ and possibly its derivatives up to some order depending on the theory) is fixed on the boundary. In order to cancel the extra, unwanted surface terms, one needs to add a so-called Gibbons-Hawking (GH) term in the action \cite{Gibbons&Hawking}. The appropriate modification of the Einstein-Hilbert action turns out to be
\begin{equation}
S^{EH}= \imj  R + 2 \isj K \label{GR_GH},
\end{equation}
where $h$ is the induced metric on the surface $\Sigma$, and $K$ is the trace of its extrinsic curvature. Variation of (\ref{GR_GH}) is then performed keeping only the metric $g_{\alpha \beta}$ fixed on the boundary.

In the original $f(R)$ action (\ref{f(R)_action}) there is no natural GH term which cancels the extra unwanted higher derivative boundary terms, except for the particular case of maximally symmetric spacetimes \cite{Bar_GH}. However, in the Jordan frame of $f(R)$ one can find an appropriate Gibbons--Hawking term in full generality, and then under the assumption that the dynamical equivalence between different representations holds on the boundary surface, one is able to re-express it in the original $f(R)$ representation. Let us examine this more carefully.

First we want to find the GH term in the Jordan frame, and so we vary (\ref{f(R)HJ}) with respect to the bulk metric and after discarding the bulk contributions we get
\begin{equation}
\delta S^{\rm{J}}_{\Sigma}=-\isj  \nabla_{\rho}  \left( \Phi \, g^{\alpha \beta} \delta \Gamma^{\rho}_{\alpha \beta} \right) =  2\isj \Phi \delta K.
\end{equation}
Therefore, the GH term that should be added in the Jordan frame action (\ref{f(R)GH}) is 
\begin{equation} 
S^{\rm{J}} _{\Sigma}= -2\isj \Phi K, \label{HJ_GH}
\end{equation}
and variation should be performed with ($\delta g_{\alpha \beta}, \delta \Phi$) vanishing on $\Sigma$.
After using the correspondence between Jordan and original frame, $\Phi \leftrightarrow f'(R)$, we find the GH term in the original frame to be \footnote{We obtained the GH term in the original $f(R)$ representation by substituting the equation of motion for $\Phi$, $V'(\Phi) =R(g)$, into Eq. (\ref{HJ_GH}).  This makes clear that the equivalence is demonstrably valid only on-shell (i.e. at the level of the classical equations of motion). }
\begin{equation} 
S^{\rm{J}} _{\Sigma}= -2\isj f'(R) K. \label{f(R)GH}
\end{equation}
However, variation of this boundary term generates a new term 
\begin{equation} 
\delta S_{\Sigma}= -2\isj f''(R) K \delta R, \label{eq:df(R)GH}
\end{equation}
which vanishes only by requiring that $\delta R = 0$ on the boundary. 
In Ref. \cite{CTHinterbichler} it has been shown that the GH term (\ref{f(R)GH}) is necessary in order to derive the correct Wald entropy for $f(R)$ gravity. However, keeping $R$ fixed on the boundary surface can be problematic. Since R includes both the first and the second derivatives of the metric, keeping it fixed would in principle require that the second derivatives of the metric are held fixed too, which overconstrains the actual formulation. The only possibility that would prevent the latter from happening would be that the condition $\delta R = 0$ is satisfied through some special configuration of the field variations on the boundary, something that would restrict the generality of our variational principle.

Therefore, we see that equivalence between the two representations breaks down at the boundary, when the consistency of the variational principle is considered.
This failure indicates that the two theories cannot be truly equivalent. Furthermore, the two theories are inequivalent at the quantum level as well; considering the path integral defined in the Jordan frame, the integration over $\Phi$, will generate extra terms in the effective action, making the latter inequivalent to the one defined in the original representation. 


\section{Dynamical equivalence of $f(R)$ gravity, part \Rmnum{2}} \label{f(R)2}
We now want to exploit the dynamical equivalence between the full $f(R)$ and Einstein--Hilbert action, using solely the Legendre transformation approach, presented in section \ref{LagrangianFormulation}, which is more general and for the $f(R)$ case gives the same result with the conformal transformation. 
This was done in Ref. \cite{Magnano1} for the bulk Lagrangian, discarding a total derivative. In the following we will show how to cure this by Legendre transforming the GH term, apart from the bulk part, getting this way the correct Einstein--Hilbert GH term as well. \footnote{The transition to the Einstein frame by means of a conformal transformation, including the relevant GH terms, has been studied in Ref. \cite{CTHinterbichler}.}

Let us begin with the action
\be
S = \imj f(R) + \isj f'(R)K,
\ee
including the GH term found in the previous section\footnote{As it was previously discussed, the GH term (\ref{f(R)GH}) can be in general problematic, however it turns to be neccessary here in order to cancel extra surface terms after Legendre transforming the action.}. As we will see below, the inclusion of the latter is indeed a good choice. We will need separate variables for the boundary surface, and the Legendre transfomed action will be of the form
\be
\widetilde{S} = \imj \left( \dot{q}_{B}p_{B} - H_{B} \right) + \isj \left( \dot{q}_{\Sigma}p_{\Sigma} - H_{\Sigma} \right), \label{f(R)HelmholtzAction}
\ee
with $H_{i} \equiv H_{i}(q,p)$.
Let us first naively associate for the generalised bulk velocity, $\dot{q} \leftrightarrow R_{\alpha \beta}$. Then we get the bulk conjugate momentum as
\be
p_{B}(q,\dot{q}) \leftrightarrow \tilde{g}^{\alpha \beta} \equiv \frac{1}{ \sqrt{-g} } \frac{\partial L } {\partial R_{\alpha \beta}} = f'(R) g^{\alpha \beta}. \label{f(R)ConjugateMomentum0}
\ee
We see that the definition of the conjugate momentum defines a conformal relation between two different metrics. In fact, as we will see below, $\tilde{g}_{\alpha \beta}$ is the metric in the Einstein frame. However, the correct association for $\dot{q}$ is not exactly $R_{\alpha \beta}$, but $R$, since relation (\ref{f(R)ConjugateMomentum0}) cannot be inverted for $R_{\alpha \beta}$. 

We proceed by identifying
\be
\dot{q}_{B} \leftrightarrow R ,\; \; \; \dot{q}_{\Sigma} \leftrightarrow K, \label{f(R)associations}
\ee
\begin{eqnarray}
& p_{B}(q, \dot{q}) \leftrightarrow \Phi \equiv \frac{1}{\sqrt{-g}} \frac{\partial L_{B}}{\partial R(g)} = f'(R), \label{f(R)BulkConjugateMomentum} \\
& p_{\Sigma}(q, \dot{q}) \leftrightarrow \Phi \equiv \frac{1}{\sqrt{-h}} \frac{\partial L_{\Sigma}}{\partial K(h)} = f'(R). \label{f(R)SurfConjugateMomentum}
\end{eqnarray}
Now, using the intuition gained from (\ref{f(R)ConjugateMomentum0}), and using (\ref{f(R)BulkConjugateMomentum})-(\ref{f(R)SurfConjugateMomentum}), we define the following relations for the bulk and surface metric respectively
\be
\tilde{g}^{\alpha \beta}  \equiv \Phi^{\frac{2}{(2-n)}}g^{\alpha \beta} \; \; \; {\rm{and}} \; \; \;  \tilde{h}^{\alpha \beta}  \equiv \Phi^{\frac{2}{(2-n)}}h^{\alpha \beta}. 
\ee

The invertibility condition is not satisfied on the boundary $\Sigma$, since $\partial^2 L_{\Sigma}/\partial K^2 =0$. However, the surface part of $\widetilde{L}$ can be still defined, with the only difference that the surface Hamiltonian will vanish identically, $H_{\Sigma} = 0$. The bulk Hamiltonian is calculated after solving one of relations  (\ref{f(R)BulkConjugateMomentum}) for $R$, $H_{B}(\Phi) \equiv \Phi f'^{-1}(\Phi)-f(f'^{-1}(\Phi))$. Using this together with (\ref{f(R)associations})-(\ref{f(R)SurfConjugateMomentum}), and substituting in (\ref{f(R)HelmholtzAction}) we arrive at

\be
\widetilde{S} =  \imj \left[  \Phi R(g) - H_{B}(\Phi)  \right] +  \isj \Phi K.     \label{f(R)LHelmholtz1}
\ee

For the transition to the Einstein frame we will use the general equations (\ref{RicciRicciDifference}) and (\ref{ExtrCurvDifference}) relating two Ricci (extrinsic curvature) tensors, evaluated for two different metrics $g_{\alpha \beta}$ ($h_{\alpha \beta}$) and $\tilde{g}_{\alpha \beta}$ ($\tilde{h}_{\alpha \beta}$).  
Defining $\Phi = \exp[\phi]$ and $\omega(n) \equiv (n-1)/(n-2)$ we get 
\begin{eqnarray}
\widetilde{S}_{B} = 
&  = \ime \left[ \widetilde{R}(\tilde{g}) - \omega(n) \pdkappa \phi \pukappa \phi - \e^{ \frac{n}{(2-n)} \phi}H_{B}(\phi) \right] \nonumber\\
&  + 2 \omega(n) \ise (\pukappa \phi) \tilde{n}_{\kappa}, \label{f(R)LHelmholtz2Bulk}
\end{eqnarray}
\begin{eqnarray}
\widetilde{S}_{\Sigma}=
&  \ise \left[ 2\widetilde{K}(\tilde{h}) - 2 \omega(n)(\pukappa \phi) \tilde{n}_{\kappa}  \right], \label{f(R)LHelmholtz2Surface}
\end{eqnarray}
and after summing up we arrive at 
\begin{eqnarray}
\hspace{-2cm} \widetilde{S} =\ime \left[ \widetilde{R}(\tilde{g}) - \omega(n) \pdkappa \phi \pukappa \phi - \e^{\frac{n}{(2-n)}\phi} H_{B}(\phi) \right] + 2\ise \widetilde{K}(\tilde{h}),\label{f(R)Einsteinframe2}
\end{eqnarray}
$\tilde{n}_{\kappa}$ denoting the normal vector to $\Sigma$. We see that we arrive at the correct, full Einstein--Hilbert action, following a conceptually different and more fundamental procedure. The conformal relation between the two metrics was revealed naturally through the definitions of the conjugate momenta. 

\section{Dynamical equivalence of $f(G)$ gravity, part \Rmnum{1}} \label{f(G)1}
In this section we will aim to express the Jordan (scalar--tensor) frame of the $R+f(G)$ action as a minimally--coupled theory by means of a conformal transformation. Firstly we will derive the GH term in the Jordan frame, and then find the appropriate one in the original frame, as dictated by the equivalence between frames. Then, in subsection \ref{f(G)2.2} we will continue with conformally transforming the full, Jordan $R+f(G)$ action.

Our starting point is the $R+f(G)$ action
\begin{equation}
S =  \imj \left[ \alpha R + f(G) \right] , \label{g(G)Lagrangian}
\end{equation}
with $G$ defined in (\ref{GB}) and $\alpha$ a dimensionless constant. 

Through the introduction of an auxiliary scalar field $\psi$ we get the Jordan frame action as
\begin{equation}
S^{\rm{J}} =  \imj \left[ \alpha R + \Phi G - V(\Phi) \right], \label{g(G)HJ}
\end{equation}
with $\Phi = f'(\psi)$, and $V(\Phi) \equiv \left[\Phi f'^{-1}(\Phi) - f \left(  f'^{-1}(\Phi) \right) \right]$, assuming that $f''(\psi) \neq 0$.

.

\subsection{The $f(G)$ Gibbons--Hawking term in the Jordan frame} \label{Sec:f(G)GH}
The motivation of this subsection is the same as in the $f(R)$ case, as explained in Section \ref{f(R)1.1}. We will derive the appropriate GH term in the original action (\ref{g(G)Lagrangian}) as dictated by the equivalence with the Jordan frame, by first calculating the Jordan frame one, presenting the explicit results of the surface parts of the action variation. Some useful variation formulas and definitions used can be found in \ref{AFormulas}.

We start from the Jordan frame action (\ref{g(G)HJ}) and vary each of the Gauss--Bonnet terms separately with respect to $g_{\alpha \beta}$ using relations (\ref{Boundary_metric})-- (\ref{Kvar}). We focus on the $f(G)$ term, since the GH term for $R$ is given by (\ref{f(R)GH}) for $f'(R)=1$.
We will again present only the boundary part of the variation, as well as work in Riemann and Gaussian normal coordinates \cite{Wheeler}. With the aid of integration by parts, and using equation (\ref{Boundary_metric}), we get
\begin{eqnarray}
&\delta S_{1\Sigma}^{\rm{J}}=- 4 \isj \, \Phi R \delta K, \\
&\delta S_{2\Sigma}^{\rm{J}}=-4  \isj  \, \Phi\bigg[ 2n^{\beta} R^{\alpha \kappa}\ndkappa-n^{\lambda}R^{\alpha \beta}  \ndlambda-n_{\lambda}h^{\alpha \beta} R^{\kappa \lambda}\ndkappa,\nonumber \\
&\;\;\;\;\;\;\;\;\;\;\;\;\;\;\;\;\;\;-n^{\alpha}n^{\beta}n_{\kappa}R^{\kappa \lambda }\ndlambda \bigg] \delta g_{\alpha \beta},\\
&\delta S_{3\Sigma}^{\rm{J}}= \isj  \, \Phi\left[n_{\lambda}R^{\alpha \kappa \lambda \beta}\ndkappa \right] \delta g_{\alpha \beta}.
\end{eqnarray}
The geometric relevance of the above terms becomes evident if we express them in terms of tensor objects defined on the boundary surface using the Gauss--Codacci equations \cite{Wheeler}. Doing this, and adding up all three terms together, we arrive at
\begin{eqnarray}
\delta S_{ \Sigma}^{\rm{J}}=\int_{\Sigma}& \; d^{n-1}x \sqrt{-h} \, \Phi \bigg[2 
\bigg( 2 \widehat G^{\beta
\gamma}\de K_{\beta \gamma}+ 2 K_{\mu}{}^{\beta}K^{\mu \gamma}\de K_{\beta \gamma} \nonumber\\
&-2KK^{\beta \gamma}\de K_{\beta \gamma} +K^2\de K -K_{\alpha
\mu}K^{\alpha \mu} \de K \bigg) \bigg], \label{g(G)STVar} 
\end{eqnarray}
with $\widehat{G}^{\beta \gamma}$ the Einstein tensor defined on $\Sigma$. 
Since we require that $\delta g_{\alpha \beta}=0$ on $\Sigma$ (or $\delta h_{\alpha \beta} =0$), it follows that $\delta \widehat{G}_{\alpha \beta}=0$
and $\delta K_{\alpha \beta}=\delta K^{\alpha \beta}=\delta K^{\alpha}{}_{\beta}$ on $\Sigma$ as well.
Using those facts, we can go backwards in (\ref{g(G)STVar}) and check that it is the variation of the following quantity
\begin{eqnarray}
S_{\Sigma}^{\rm{J}}  = \isj&  \,  \Phi \left[2\left( 2 \widehat G^{\alpha \beta} K_{\alpha \beta}+ J \right) \right], \label{g(G)GH} 
\end{eqnarray}
with $J \equiv \frac{2}{3}K^{\rho}{}_{\kappa}K^{\kappa \lambda}K_{\lambda \rho}-KK_{\kappa \lambda}K^{\kappa \lambda} +\frac{1}{3}K^{3}$. The appropriate supplement for the initial scalar--tensor action is therefore equation (\ref{g(G)GH}) with a minus sign instead. The GH term for a simple Gauss--Bonnet action ($L \propto \sqrt{-g}G$) has been derived under more general assumptions in a braneworld context in Ref. \cite{Davis}, as well as in Ref. \cite{Meyers} using the calculus of differential forms.

Now, as in the $f(R)$ case, we can use the equivalence $f'(G) \leftrightarrow \Phi$, to find the GH term in the original $f(G)$ frame if the equivalence is to hold on the boundary,
\begin{equation}
S_{\Sigma}= -\isj  \, f'(G) \left[2\left( 2 \widehat G^{\alpha \beta} K_{\alpha \beta}+ J \right) \right]. \label{g(G)HJAction} 
\end{equation}
Now, variation of the action requires $\delta g_{\alpha \beta} =0$ and $\delta G = 0$ on $\Sigma$. The latter condition can yield a problematic variational principle for the same reasons discussed in Section \ref{f(R)1.1}. Therefore, for $R+f(G)$ theories as well, true equivalence on the boundary surface is broken.

\subsection{Conformal transformation of the Jordan frame action} \label{f(G)2.2}
We now want to study if the non--minimally coupled, full Jordan frame action (\ref{g(G)HJAction}), can be decoupled from $\Phi$ and written in a $G +$ scalar field form, similar to the $f(R)$ case, using a conformal transformation of the metric.

Before we start with the calculations, let us introduce a notation that will make our equations look shorter. So, only for the rest of this section, we shall define: $n_{i} \equiv (n-i)$ and $r_{i} \equiv 1/n_i$, where $n$ is the spacetime dimensionality. $n_i$ is not to be confused with the surface normal $n_{\alpha}$.

We shall begin with the bulk term. Using the transformation formula (\ref{CTGB}) together with the conformal factor identification
\begin{equation}						
\Omega = \Phi^{1/(n-4)}\equiv \Phi^{r_{4}},  \label{g(G)CTI}
\end{equation}
and omitting the potential which transforms trivially, the action (\ref{g(G)HJ}) after the redefinition $\Phi = \exp[\phi]$ becomes
\begin{eqnarray}
&\hspace*{-2cm}\imj \, \Phi G \mapsto \ime \Bigg\{ \widetilde{G} \no\\
& - 8r_{4}n_{3}\left[ \phi^{\tilde{;}\alpha \beta} - r_{4}n_{5} \phi^{,\alpha}  \phi^{,\beta}  \right]\widetilde{R}_{\alpha \beta} 
-2r_{4}n_{3} \left[ 3r_{4}n_{4} \phi^{,\kappa}  \phi_{,\kappa}  - 2\widetilde{\Box} \phi - ae^{ \left(1 - r_{4}n_{2}\right) \phi } \right] \widetilde{R}\no \\
& + 4r_{4}^{4}n_{3}n_{2} \left[ 2n_{5}n_{3} + \frac{nn_{1}}{4} - 3a r_{4}^{2}n_{5}n_{1} e^{ \left(1-r_{4}n_{2}\right) \phi }  \right]  \left( \phi_{,\kappa}  \phi^{,\kappa}  \right)^2  \no \\
& + 4r_{4}^{3}n_{5}n_{3}n_{2} \bigg[ \phi_{,\alpha}  \phi_{,\beta}  + 2\phi_{\tilde{;}\alpha \beta}   \bigg]  \left( \phi^{,\alpha}  \phi^{,\beta}  \right)  \no \\
& + 4r_{4}^{2} n_{3}n_{2} \bigg[\widetilde{\Box} \phi - r_{4}\left( 2n_{5}+n_{1} \right)  \phi^{,\kappa}  \phi_{,\kappa} +  2ar_{4}n_{1}e^{ \left( 1 -r_{4}n_{2}\right) \phi } \bigg] \widetilde{\Box} \phi \no\\
& - 4r_{4}^{2}n_{3}n_{2} \phi^{\tilde{;}\alpha \beta} \;  \phi_{\tilde{;}\alpha \beta} 
\Bigg\}, \label{g(G)CT2}
\end{eqnarray}
with $\widetilde{\Box} \equiv \widetilde{\nabla}^{\kappa}\widetilde{\nabla}_{\kappa}$.

Identification (\ref{g(G)CTI}) breaks down for $n = 4$, and in fact it is valid only for $n \geq 5$. This means that we are unable to decouple the scalar $\Phi$ from the Gauss--Bonnet term unless $n \geq 5$. For $n \geq 5$, the GB term is minimally coupled to the scalar $\Phi = \rm{exp}[\phi]$, but there are new couplings between the derivatives of $\phi$, the Ricci tensor and Ricci scalar. In this case, action (\ref{g(G)CT2}), plus the scalar potential term of (\ref{g(G)HJ}), describes a fourth order, non minimally coupled scalar--tensor theory. 

Let us now turn attention to the conformal transformation of the relevant GH term, calculated in \ref{ASupplementGBGH}. One can see that the variational principle requires that we impose apart from $\delta \tilde{g} =0$ and  $\delta \phi =0$, the extra conditions $ \widetilde{\nabla}\delta \tilde{g}=0$ and $\widetilde{\nabla}\delta \phi=0$ on $\Sigma$. The $R+f(G)$ action cannot be expressed as a second order, minimally--coupled scalar tensor one, in contrast with $f(R)$ gravity.

\section{Dynamical equivalence of $f(G)$ gravity, part \Rmnum{2}} \label{f(G)2}
The richer structure of the $R+f(G)$ action gives us more options in identifying generalised velocities, compared to the $f(R)$ one. In this section we want to take advantage of the latter fact, and re-express the original $R+f(G)$ action as a new one with not only a new scalar, but with new tensor fields as well, by means of a Legendre transformation. The new variational principle will be of second order.

Our starting point is the action
\be
S = \imj \left[ \alpha R+f(G)  \right].
\ee
We proceed with defining our conjugate momenta as 
\begin{eqnarray}
& p_1  \leftrightarrow \Psi \equiv \frac{1}{\sqrt{-g}}\frac{\partial L}{\partial R} = \alpha + 2Rf'(G), \label{f(G)2momentum1}\\
& p_2 \leftrightarrow \tilde{g}^{\alpha \beta} \equiv \frac{1}{\sqrt{-g}}\frac{\partial L}{\partial R_{\alpha \beta}} = -8f'(G)R^{\alpha \beta}, \label{f(G)2momentum2}\\
&p_3  \leftrightarrow \sigma_{\alpha}{}^{\beta \gamma \delta} \equiv \frac{1}{\sqrt{-g}}\frac{\partial L}{\partial R^{\alpha}{}_{\beta \gamma \delta}} = 2 f'(G)R_{\alpha}{}^{\beta \gamma \delta}\label{f(G)2momentum3}. 
\end{eqnarray}

Defining $\Phi \equiv \Phi(G) \equiv f'(G)$, the inverse of the above relations read
\begin{eqnarray}
&R = \frac{1}{2\Phi} \left(\Psi - \alpha  \right), \label{f(G)2momentum1Inv}\\
&R^{\alpha \beta} =-\frac{1}{8\Phi} \tilde{g}^{\alpha \beta} , \label{f(G)2momentum2Inv}\\
&R_{\alpha }{}^{\beta \gamma \delta}(\Phi,g,\tilde{g},\sigma) = \frac{1}{2\Phi} \sigma_{\alpha}{}^{\beta \gamma \delta}, \label{f(G)2momentum2Inv}
\end{eqnarray}
with $\tilde{g}^{\kappa}{}_{\kappa} \equiv g^{\kappa \lambda} \tilde{g}_{\kappa \lambda}$ and $f'(G) \neq 0$. In fact, we will use $g_{\alpha \beta}$ to raise and lower indices for the rest of the section. 

For the calculation of the Hamiltonian we will need to express the Gauss--Bonnet term in terms of the new fields $(\Psi, \tilde{g}^{\alpha \beta}, \sigma_{\alpha}{}^{\beta \gamma \delta})$. Using the inverse relations (\ref{f(G)2momentum1Inv})-(\ref{f(G)2momentum2Inv}) we get
\begin{eqnarray}
G &\equiv R^2 -4R^{\mu \nu} R_{\mu \nu} + R^{\mu}{}_{\nu \rho \sigma} R_{\mu}{}^{\nu \rho \sigma} = \frac{1}{4\Phi^2} \Gamma(\Psi,\tilde{g},\sigma), \label{GB2}
\end{eqnarray}
with the function $\Gamma$ defined as
\be
\Gamma(\Psi,\tilde{g},\sigma) \equiv \left( \Psi - \alpha \right)^2 - \frac{\tilde{g}^{\mu \nu} \tilde{g}_{\mu \nu}}{4} + \ \sigma_{\mu}{}^{\nu \rho \sigma}  \sigma^{\mu}{}_{\nu \rho \sigma}.
\ee
Furthermore, we assume that we can invert relation (\ref{GB2}) \footnote{The necessary condition is that $[\left(f'\left(G\right) \right)^2G]' \neq 0$, implying that $f(G) \neq C_1\sqrt{G}+C_2$.} and express the Gauss--Bonnet term in terms of the function $\Gamma$ as
\be
G = G^{-1}(\Gamma) \equiv J(\Gamma),
\ee
so that
\be
f'(G) = f'(J(\Gamma)) \equiv F(\Gamma).
\ee
Using all the above, we can now calculate the Hamiltonian as
\begin{eqnarray}
\hspace{-1.5cm }H(\Gamma(\Psi,\tilde{g}, \sigma)) &= \Psi R(\Psi,\tilde{g},\sigma) + \tilde{g}^{\alpha \beta}R_{\alpha \beta}(\Psi,\tilde{g},\sigma) +  \sigma_{\alpha}{}^{\beta \gamma \delta} R^{\alpha}{}_{\beta \gamma \delta}(\Psi,\tilde{g},\sigma) - (-g)^{-1/2}L(\Psi,\tilde{g}, \sigma) \nonumber \\
&=  \frac{\Gamma}{2F(\Gamma)} - f\left( J(\Gamma) \right).
\end{eqnarray}
Notice that the fields $(\Psi, \tilde{g}^{\alpha \beta}, \sigma_{\alpha}{}^{\beta \gamma \delta})$ enter implicitly in the Hamiltonian through the function $\Gamma$.

The Legendre transformed action then reads
\begin{eqnarray}
\hspace{-2.7cm}\widetilde{S}[\Psi,g,\tilde{g},\sigma] = \imj &\Bigg[ \Psi R(g) + \tilde{g}^{\alpha \beta} R_{\alpha \beta}(g) + \sigma_{\alpha}{}^{\beta \gamma \delta} R^{\alpha}{}_{\beta \gamma \delta}(g) - H(\Gamma(\Psi,\tilde{g}, \sigma))\Bigg]. \ \label{g(G)LHelmholtz}
\end{eqnarray}
To get the equations of motion we vary the action $\widetilde{S}$ with respect to the four fields $(\Psi, \tilde{g}^{\alpha \beta}, \sigma_{\alpha}{}^{\beta \gamma \delta},g_{\alpha \beta})$ to get
\begin{eqnarray}
&\frac{\delta \widetilde{S} }{\delta \Psi} =  R(g) - 2 H'\left( \Psi - \alpha  \right)   = 0, \label{f(G)2eom1} \\ \nonumber \\ 
&\frac{ \delta \widetilde{S} }{ \delta \tilde{g}^{\alpha \beta} } = R_{\alpha \beta}(g)  + \frac{1}{2}H' \tilde{g}_{\alpha \beta}  = 0, \label{f(G)2eom2} \\ \nonumber \\ 
&\frac{ \delta \widetilde{S} }{ \delta \sigma_{\alpha}{}^{\beta \gamma \delta} } =  R^{\alpha}{}_{\beta \gamma \delta}(g) - 2H' \sigma^{\alpha}{}_{\beta \gamma \delta} =0,  \label{f(G)2eom3}
\end{eqnarray}

\begin{eqnarray}
\frac{ \delta \widetilde{S} }{ \delta g_{\alpha \beta} } &=   \Psi G^{\alpha \beta} - \nualpha \nubeta \Psi + g^{\alpha \beta} \nukappa \ndkappa \Psi  - \ndkappa \nabla^{(\alpha} \tilde{g}^{\beta) \kappa} \nonumber  \\
&  +  \frac{1}{2} \nurho \ndrho \tilde{g}^{\alpha \beta} + \frac{1}{2}g^{\alpha \beta} \ndkappa \ndlambda \tilde{g}^{\kappa \lambda} - 2 \ndkappa \ndlambda \sigma^{\kappa (\alpha  \beta) \lambda}\nonumber  \\
&   - \frac{1}{2} g^{\alpha \beta} \Big[ \tilde{g}^{\kappa \lambda} R_{\kappa \lambda}(g) + \sigma_{\kappa}{}^{\lambda \mu \nu} R^{\kappa}{}_{\lambda \mu \nu}(g) - H(\Gamma)\Big] \nonumber \\
&- \frac{1}{2} H' \Big[ 8 \sigma^{ \kappa \lambda \mu (\alpha}\sigma^{\beta)}{}_{\mu \lambda \kappa} -  \tilde{g}^{\kappa (\alpha}\tilde{g}_{\kappa}{}^{ \beta)}  \Big]= 0, \label{f(G)2eom4}
\end{eqnarray} 
with $G^{\alpha \beta} \equiv R^{\alpha \beta} - \frac{1}{2}g^{\alpha \beta}R$, $H' \equiv H'(\Gamma) \equiv \partial H/\partial \Gamma$ and covariant derivatives $\nabla_{\alpha}$ defined with respect to $g_{\alpha \beta}$.

Variation with respect to $g_{\alpha \beta}$ yields surface terms $\propto \nabla g_{\alpha \beta}$. We want to keep only the fields fixed on $\Sigma$ and not their derivatives, so we have to add in action (\ref{g(G)LHelmholtz}) the following GH term
\begin{eqnarray}
\widetilde{S}_{\Sigma} =   2\isj  \, \Big ( \Psi K + \tilde{g}^{\alpha \beta} \Gamma^{\kappa}_{\alpha [\kappa}n_{\beta]} + \sigma_{\alpha}{}^{\beta \gamma \delta} \Gamma^{\alpha}_{\beta [\gamma}n_{\delta]} \Big), \label{GHf(G)2}
\end{eqnarray}
with $[A,B] \equiv \frac{1}{2}\left( AB - BA \right)$.

If we now contract equation (\ref{f(G)2eom3}) with $g_{\alpha}{}^{\gamma}$ and add it to (\ref{f(G)2eom3}) we get the relation
\be
\sigma^{\gamma}{}_{\alpha \gamma \beta} \equiv \sigma_{\alpha \beta} = -\frac{1}{4} \tilde{g}_{\alpha \beta} \label{Sigma_Tildeg_Relation}.
\ee  
The latter implies that $\sigma_{\alpha \beta \gamma \delta}$ can be expressed as some combination of $g_{\alpha \beta}$ and $\tilde{g}_{\alpha \beta}$ plus some traceless part, while the trace of that expression should give (\ref{Sigma_Tildeg_Relation}). To find the latter expression we can expand the Riemann tensor in terms of he Ricci tensor and scalar according to 
\be
R_{\alpha \beta \gamma \delta} = C_{\alpha \beta \gamma \delta} - a_{n} \left(  g_{\alpha [\delta}R_{\gamma] \beta} + g_{\beta [\gamma}R_{\delta] \alpha} \right) - b_{n} R g_{\alpha [ \gamma}g_{\delta] \beta}, \label{Riemann_Expansion }
\ee
with $a_{n} \equiv  \frac{2}{n-2} $, $b_{n} \equiv  \frac{2}{(n-1)(n-2)}$ and $C_{\alpha \beta \gamma \delta}$ the Weyl tensor which is traceless in all its indices. After use of equation (\ref{f(G)2eom2}), relation (\ref{Riemann_Expansion }) can be expressed as
\be
R_{\alpha \beta \gamma \delta} = C_{\alpha \beta \gamma \delta} + \frac{a_{n}}{2} H' \left(  g_{\alpha [\delta}\tilde{g}_{\gamma] \beta} + g_{\beta [\gamma}\tilde{g}_{\delta] \alpha} \right) + \frac{b_{n}}{2} H' \tilde{g} g_{\alpha [ \gamma}g_{\delta] \beta},
\ee
and plugging the latter into equation (\ref{f(G)2eom3}) to substitute for the Riemann tensor, we get a relation between $\sigma_{\alpha \beta \gamma \delta}$, $g_{\alpha \beta}$ and $\tilde{g}_{\alpha \beta}$
\be
\sigma_{\alpha \beta \gamma \delta}   =  \frac{1}{2H'} C_{\alpha \beta \gamma \delta}(g) + \frac{a_{n}}{4} \left(  g_{\alpha [\delta}\tilde{g}_{\gamma] \beta} + g_{\beta [\gamma}\tilde{g}_{\delta] \alpha} \right) +  \frac{b_{n}}{4} \, \tilde{g} g_{\alpha [ \gamma}g_{\delta] \beta}. \label{sigma_Condition}
\ee

Combining equations (\ref{f(G)2eom1}) and (\ref{f(G)2eom2}) we can find a similar relation for $\Psi$
\be
\Psi(g, \tilde{g}) = \alpha - \frac{1}{4}\tilde{g}, \label{g_Condition}
\ee
with $\tilde{g} \equiv g^{\alpha \beta} \tilde{g}_{\alpha \beta}$.
One would like to be able to solve equation (\ref{sigma_Condition}) for $\sigma_{\alpha \beta \gamma \delta} = \sigma_{\alpha \beta \gamma \delta} (\tilde{g}, g)$. However, this is not in principle possible unless $H' = {\rm constant}$ (corresponding to the trivial case of $f(G) = G$) or $C_{\alpha \beta \gamma \delta}(g) = 0$. The latter case includes the case of the FRW spacetime or maximally symmetric spacetimes like the Minkowski one. In that case, all fields can be expressed in terms of $g_{\alpha \beta}$ and $\tilde{g}_{\alpha \beta}$ and we can get a solution for the latter ones by solving the appropriate system of second order differential equations, which we derive below.

Now, we want to derive a system of evolution equations for the set of fields ($g_{\alpha \beta}, \tilde{g}_{\alpha \beta}$). The first equation we will use results from equation (\ref{f(G)2eom2}) after taking its trace once, together with some simple algebra. To get the second equation, we use relations (\ref{sigma_Condition}) and (\ref{g_Condition}) together with the $C_{\alpha \beta \gamma \delta}(g) = 0$ ansatz to express the last of the equations of motion, equation (\ref{f(G)2eom4}), in terms of $g_{\alpha \beta}$ and $\tilde{g}_{\alpha \beta}$ alone. This way we arrive at the new system of second order equations for the set of fields $(g_{\alpha \beta}, \tilde{g}_{\alpha \beta})$
\be
G_{\alpha \beta} = -\frac{1}{2} H' \left( \tilde{g}_{\alpha \beta} - \frac{1}{2} g_{\alpha \beta} \tilde{g}^{\kappa}{}_{\kappa} \right) ,\label{Einstein_tensor_new}
\ee

\begin{eqnarray}
& \left( \widehat{P}^{\kappa \lambda}_{\mu \nu} \right)^{(\alpha \beta)} \nabla_{\kappa} \nabla_{\lambda} \tilde{g}^{\mu \nu}  - H' g^{\alpha \beta} \Big[ p_{n}\tilde{g}^{\kappa \lambda} \tilde{g}_{\kappa \lambda} + q_{n} \tilde{g}^{2} - \alpha \tilde{g} - 2 \frac{H}{H'} \Big] \no \\
&- H' \Big[ r_{n} \tilde{g}^{\kappa (\alpha}\tilde{g}^{\beta)}{}_{\kappa}   + s_{n} \tilde{g}\tilde{g}^{\alpha \beta} + 2\alpha \tilde{g}^{\alpha \beta} \Big]  =  0, \label{New_system_Eq_2}
\end{eqnarray}
with $H \equiv H(\Gamma(g,\tilde{g}))$ and the operator $\widehat{P} \equiv \widehat{P}(g)$ as well as the constants $p_{n}, q_{n}, r_{n}, s_{n}$ defined in \ref{ASupplementGBRep2}. To arrive at equation (\ref{New_system_Eq_2}) we have used the following relations
\be
\Gamma \equiv \Gamma(g, \tilde{g}) = \frac{1}{16} \left[  \left(  1  -  \frac{4(n-3)}{(n-2)^2 (n-1)}  \right)\tilde{g}^2  - 4 \left( \frac{ n-3}{n-2} \right) \tilde{g}^{\mu \nu}\tilde{g}_{\mu \nu}  \right] ,
\ee
and 
\be
\hspace{-1cm} \sigma^{\alpha \mu \nu \rho} \sigma^{\beta}{}_{\mu \nu \rho} (g, \tilde{g}) = \frac{1}{8(n-2)^2} \left[ g^{\alpha \beta} \tilde{g}^{\rho \mu} \tilde{g}_{\rho \mu} + (n-4)\tilde{g}^{\rho \alpha}\tilde{g}^{\beta}{}_{\rho} - 2 \left( \frac{n-3}{n-1} \right) \tilde{g}\tilde{g}^{\alpha \beta} \right],
\ee
as well as $\sigma^{\alpha}{}_{\beta \gamma \delta}(g, \tilde{g})$ given by (\ref{sigma_Condition}) with $C^{\alpha}{}_{\beta \gamma \delta}(g) = 0$. 

A look at the first equation of the new system, equation (\ref{Einstein_tensor_new}), shows that at the level of the equations of motion we can express the dynamics as GR, minimally--coupled to an effective energy--momentum tensor (the r.h.s of the equation) described by the spin two field $\tilde{g}_{\alpha \beta}$.

There is one extra constraint the fields satisfy, that is the Bianchi identities. 
Since the l.h.s of equation (\ref{Einstein_tensor_new}) is covariantly conserved, as dictated by the Bianchi identities, then the r.h.s should be as well, 
\be
\nabla^{\alpha} \Big[ H' \Big( \tilde{g}_{\alpha \beta} - \frac{1}{2} g_{\alpha \beta} \tilde{g} \Big) \Big] = 0.
\ee
The latter equation is a condition the set of fields $(g_{\alpha \beta}, \tilde{g}_{\alpha \beta})$ have to satisfy, together with the equations of motion.

We will not seek solutions of the system described by equations (\ref{Einstein_tensor_new})-(\ref{New_system_Eq_2}) in this paper, leaving this for a possible future work. However, it is easy to see that Minkowski space is a solution for ($g_{\alpha \beta}, \tilde{g}_{\alpha \beta}$) = ($\eta_{\alpha \beta}, 0$).

We see that at the classical level of this representation there are two independent fields, ($g_{\alpha \beta}, \tilde{g}_{\alpha \beta}$), satisfying a system of second order equations together with a second order condition, the Bianchi identity. These equations should be classically equivalent to the original fourth order ones, as $\tilde{g}_{\alpha \beta}$ is related to the second derivatives of $g_{\alpha \beta}$. It should be noted that there is no reason to expect complete equivalence of the quantum equations, as the measure of the path integral in the different representations can introduce new terms. 


\section{Conclusions} \label{conclusions}
Legendre transformations are a fundamental tool to study the dynamical equivalence between different modified gravity actions, with the aim of understanding better the nature of the theories under study. When working the level of the action, a consistent analysis should take into account the appropriate Gibbons--Hawking (GH) terms (full action). Although in a general context there are no natural GH terms for both $f(R)$ and $R+f(G)$ actions, however one can define them considering the dynamical equivalence between two different representations of the particular action on the boundary surface, as it was done in Section \ref{Sec:f(G)GH}, when we considered the equivalence between the original action and the Jordan frame one. However, the GH terms found through this procedure turn out to render the variational principle inconsistent. This is due to the fact that the two representations are not equivalent at the quantum level, as pointed out in Section \ref{f(R)1.1}.

Due to the structural simplicity of the full $f(R)$ action, the Legendre transformation yields in this case the same result as a conformal transformation of the original action. On the other hand, the $R+f(G)$ Jordan (non--minimally coupled) frame action cannot be re-expressed as a second order theory through a conformal transformation, despite the fact that the auxiliary scalar decouples from the Gauss--Bonnet term for $dim \geq 5$. The resulting theory is still of fourth--order, as was calculated explicitly for the full action in Section \ref{f(G)1}.

However, the more complex structure of the $R + f(G)$ action, allows one to re-express it, by means of a Legendre transformation, as a second order theory with extra tensor fields apart from scalars. In the new representation, it turns out that only two fields are the independent ones, the metric $g_{\alpha \beta}$ and the rank two field $\tilde{g}_{\alpha \beta}$. At the level of the equations of motion, we are able to recover GR, sourced by an effective energy--momentum tensor, which is a function of $\tilde{g}_{\alpha \beta}$. Although the two representations are classically equivalent in vacuum without boundary, at the quantum level they differ, as integrating out the extra fields generates new terms in the effective action.

\ack
IDS is grateful to Martin Kunz and Andrew Liddle for valuable discussions, comments and encouragement. Furthermore, he would like to thank Antonio De Felice for useful feedback on the manuscript, as well as Thomas Sotiriou for a useful discussion on boundary terms. IDS is supported by GTA funding from the University of Sussex.
\begin{appendix}

\section{Definitions and formulas} \label{AFormulas}
The $n$--dimensional bulk metric $g_{\alpha \beta}$ induces an $(n-1)$--dimensional metric $h_{\alpha \beta}$ on the boundary surface $\Sigma$ as 
\be
h_{\alpha \beta}=g_{\alpha \beta} \pm n_{\alpha}n_{\beta}, \label{Boundary_metric}
\ee
for a spacelike ($+$) and timelike ($-$) surface $\Sigma$ respectively and $n_{\alpha}$ is the normal to the surface $\Sigma$. $h_{\alpha \beta}$ can be used as a projection operator from the tangent space to the bulk $M$ to the tangent space to the boundary $\Sigma$ at a point $P_{0}$. Particularly, for its action on $n_{\alpha}$ it is $h^{\alpha \beta} n_{\beta} = 0$.

The extrinsic curvature $K_{\alpha \beta}$ is an $(n-1)$--dimensional tensor that measures the ``bending" of  $\Sigma$ in the bulk spacetime $M$, and is defined as 
\begin{eqnarray}
K_{\alpha \beta} =  \frac{1}{2}\pounds_{n}h_{\alpha \beta} = \ndalpha \xi_{\beta} = h^{\gamma}{}_{\alpha}\nabla_{\gamma} \xi_{\beta} =h_{\alpha}{}^{\gamma}\nabla_{\gamma} n_{\beta},
\end{eqnarray}
where ``$\pounds$" is the Lie derivative, $\xi^{\beta}$ a unit tangent to the geodesic congruences orthogonal to $\Sigma$, $n^{\beta}$ any other normal to $\Sigma$, and $\nabla_{a}$ defined with respect to the bulk metric $g_{\alpha \beta}$. 

The following variation formulas are used to calculate the GH terms presented in Sections \ref{f(R)1}, \ref{f(G)1} and \ref{f(G)2}. They are evaluated using the special coordinate systems of Gauss and Riemann normal coordinates respectively \cite{Wheeler}. We have
\begin{eqnarray}
&\delta \Gamma^\sigma{}_{\mu \nu}= \frac{1}{2}g^{\sigma
\rho}\left( \delta g_{\nu \left[\rho, \mu \right]}+ \delta g_{\rho \mu, \nu} \right), \\ \label{GammaVariation}
&\delta R^{\alpha}{}_{\beta \gamma \delta} = g^{\alpha \kappa} \left( \delta g_{\kappa [\delta ; \gamma] \beta}- \delta g_{\beta[ \gamma ; \delta] \kappa} \right), \label{RiemannVariation} \\ 
&\delta K_{\alpha \beta} = \; - h_{\alpha}{}^{\gamma}\delta \Gamma^{\delta}_{\beta \gamma} n_{\delta } = \; -\frac{1}{2}\,n_{\delta}\,h_{\alpha}{}^{\gamma}\, \, g^{\delta
\rho} \nabla_{[\rho}\delta g_{\beta] \gamma}, \\
&\delta K \equiv \delta K^{\alpha}{}_{\alpha} = \frac{1}{2} n^{\rho} h^{\alpha \gamma}
\nabla_{\rho} \de g_{ \alpha \gamma}, \label{Kvar}
\end{eqnarray}
with $[A,B] \equiv \frac{1}{2}\left( AB-BA \right)$.
The variation of the Ricci tensor and scalar can be found beginning from (\ref{RiemannVariation}) and calculating the variation of the appropriate contractions, for example, $\delta R_{\beta \delta} \equiv \delta (g^{\gamma}{}_{\alpha} R^{\alpha}{}_{\beta \gamma \delta})$.

If $g_{\alpha \beta} \mapsto \tilde{g}_{\alpha \beta}= \Omega^{2} g_{\alpha \beta}$ we have \cite{Wald, CTDabrowski}
\begin{eqnarray}
R = \Omega^{2}\left[\widetilde{R}+2(n-1)\Omega^{-1}\widetilde{\Box} \Omega-n(n-1)\Omega^{-2}\tilde{g}^{\alpha \beta}\Omega_{,\alpha}\Omega_{,\beta} \right], \label{CTRicciScalar}
\end{eqnarray}
\begin{eqnarray}
G = &\Omega^{4}  \bigg[ \widetilde{G}-4n_{3}\Omega^{-1}\left( 2\widetilde{R}_{\alpha \beta} \Omega^{\tilde{;}\alpha \beta}-\widetilde{R} \widetilde{\Box} \Omega \right) \nonumber \\
&+2n_{2}n_{3}\Omega^{-2} \left( 2(\widetilde{\Box}\Omega)^2-2\Omega_{\tilde{;}\alpha \beta} \Omega^{\tilde{;}\alpha \beta} - \widetilde{R} \Omega_{,\kappa} \Omega^{,\kappa} \right) \nonumber \\
&-n_{1}n_{2}n_{3}\Omega^{-3}\left( 4 (\widetilde{\Box} \Omega)\Omega_{,\kappa} \Omega^{, \kappa} - n\Omega^{-1}(\Omega_{,\kappa} \Omega^{, \kappa})^{2} \right)
\bigg], \label{CTGB}
\end{eqnarray}
\be
K_{\alpha \beta}=\Omega^{-1} \left[\widetilde{K}_{\alpha \beta}-\Omega^{-1} \tilde{h}_{\alpha \beta} \Omega_{,\kappa} \tilde{n}^{\kappa}  \right], \label{Kab_conformal}
\ee 
where in (\ref{CTGB}) we use the convention $n_{i} \equiv (n-i)$, with $n$ the spacetime dimension.

For two different metrics $g_{\alpha \beta}$ and $\tilde{g}_{\alpha \beta}$, defined on the same manifold $M$, and not necessarily conformally related, we have
\begin{equation}
\hspace{-1.5cm} R_{\alpha \beta \gamma}{}^{\delta}(g) - \widetilde{R}_{\alpha \beta \gamma}{}^{\delta}(\tilde{g}) = \widetilde{\nabla}_{\beta} C^{\delta}{}_{\alpha \gamma}- \widetilde{\nabla}_{\alpha} C^{\delta}{}_{\beta \gamma} + C^{\kappa}{}_{\alpha \gamma}C^{\delta}{}_{\beta \kappa}-C^{\kappa}{}_{\beta \gamma}C^{\delta}{}_{\alpha \kappa}, \label{RicciRicciDifference}
\end{equation}
\begin{eqnarray}
K_{\alpha \beta}(h) - \widetilde{K}_{\alpha \beta}(\tilde{h}) &= \frac{1}{2}\pounds_{n}\tilde{h}_{\alpha \beta} - \frac{1}{2}\pounds_{n} h_{\alpha \beta}  = -h_{\alpha}{}^{\gamma}C^{\kappa}{}_{\gamma \beta}n_{\kappa}, \label{ExtrCurvDifference}
\end{eqnarray}
with $C^{\alpha}{}_{\beta \gamma} \equiv \frac{1}{2}g^{\alpha \sigma}( \widetilde{\nabla}_{\beta} g_{\gamma \sigma}+\widetilde{\nabla}_{\gamma} g_{\beta \sigma}-\widetilde{\nabla}_{\sigma} g_{\beta \gamma}) $. 
Particularly in Section \ref{f(R)2}, where $\tilde{g}_{\alpha \beta} = \Phi g_{\alpha \beta} \; (\Omega^2 \equiv \Phi)$, equations (\ref{RicciRicciDifference}) and (\ref{ExtrCurvDifference}) are used in the form below
\begin{equation}
\hspace{-1cm} R(g) = \Phi \left[ \widetilde{R}(\tilde{g}) - \frac{1}{4}(n-2)(n-1)\Phi^{-2}\pdkappa \Phi \pukappa \Phi + (n-1)\tnukappa(\Phi^{-1}\partial_{\kappa} \Phi) \right], 
\end{equation}
\begin{eqnarray}
K \equiv h^{\alpha \beta}K_{\alpha \beta} &= \Phi^{1/2} \left[ \widetilde{K} - \frac{1}{2} (n-1) \tilde{n}^{\kappa}  \Phi^{-1} \partial_{\kappa} \Phi \right], 
\end{eqnarray} 
with $\tilde{n}_{\alpha} = \Phi^{1/2}n_{\alpha}$ and $ \widetilde{K}  \equiv \tilde{h}^{\alpha \beta} \tndalpha \tilde{n}_{\beta}$.

\section{Conformal transformation of the Gauss--Bonnet GH term} \label{ASupplementGBGH}
Here we will present the conformal transformation of the Gauss--Bonnet GH term, (\ref{g(G)GH}). For the two terms of (\ref{g(G)GH}) we get respectively
\begin{eqnarray}
\hspace*{-2cm}&\isj\; J \mapsto \ise \; \Omega^{4-n} \;\Phi \Bigg\{ \widetilde{J} \no\\
& + n_{3}\Omega^{-1} \left[ \widetilde{K}^2 - \widetilde{K}_{\alpha \beta}\widetilde{K}^{\alpha \beta} \right] \left( \Omega_{,\kappa} \tilde{n}^{\kappa} \right) 
- n_{3}n_{2}  \widetilde{K}\; \Omega ^{-2}\left( \Omega_{,\kappa} \tilde{n}^{\kappa}  \right)^{2}  \no \\
& + \frac{1}{3} n_{3}n_{2}n_{1} \Omega^{-3} \left( \Omega_{,\kappa} \tilde{n}^{\kappa}  \right)^{3}
\Bigg\}.
\label{GBGH11}
\end{eqnarray}
\begin{eqnarray}
\hspace*{-2cm}& \isj \;  \Phi \widehat{G}^{\alpha \beta} K_{\alpha \beta} \mapsto  \ise \; \Omega^{4-n}  \Phi \Bigg\{ \widetilde{ \widehat{G}}^{\alpha \beta}\widetilde{K}_{\alpha \beta} \no \\
&  +  n_{3} \Omega^{-1} \left[  \frac{1}{2} \widetilde{\widehat{R}} (\Omega_{,\kappa} \tilde{n}^{\kappa})+ \widetilde{K}_{\alpha \beta}\Omega^{\tilde{;}\alpha \beta}  - \widetilde{K}\widetilde{\Box}\Omega  \right] \no \\
&  +  n_{3}\Omega^{-2}\left[ \frac{n_{2}}{2} \widetilde{K} (\Omega_{,\kappa}  \Omega^{,\kappa} )- \widetilde{\Box}\Omega( \Omega_{, \kappa}  \tilde{n}^{\kappa})+n_{1}\widetilde{\Box} \Omega (\Omega_{,\kappa}  \tilde{n}^{\kappa} ) \right] \no \\
& - \frac{1}{2} \Omega^{-3} n_{3}n_{2}n_{1} \left( \Omega_{,\kappa}  \Omega^{,\kappa}\right)^{2} \left( \Omega_{,\lambda} \tilde{n}^{\lambda} \right) \Bigg \}.
\label{GBGH12}
\end{eqnarray}

Adding up terms (\ref{GBGH11}) and (\ref{GBGH12}) we get the GB GH term in the conformally transformed frame. However, boundary terms resulting by variation of action (\ref{g(G)CT2}) with respect to $\tilde{g}_{\alpha \beta}$ and $\Phi$, will not be able to cancel with the GH subterms in (\ref{GBGH11}) and (\ref{GBGH12}), as was the case in $f(R)$. Consequently, we are left with terms proportional to first and second order derivatives of both the metric and scalar field on the boundary surface $\Sigma$, which should be held fixed in the initial value formulation, together with $g_{\alpha \beta}$ and $\phi$ themselves, in order for the GH term to be zero in the total variation.


\section{Complementary definitions for Section \ref{f(G)2} } \label{ASupplementGBRep2}
Here we define the operator and the constants used in equation (\ref{New_system_Eq_2}). The operator is defined as
\begin{eqnarray}
\hspace{-1.5cm} \left( \widehat{P}^{\kappa \lambda}_{\mu \nu}\right)^{(\alpha \beta)}   \equiv   (1 - a_{n}) \Big[ & c_{n} g^{\kappa (\alpha} g^{\beta) \lambda} g_{\mu \nu}   - c_{n}  g^{\alpha \beta} g^{\kappa \lambda}g_{\mu \nu}   -  4 g^{\lambda ( \alpha}\delta^{\beta)}_{\mu} \delta^{\kappa}_{\nu}  +  2 g^{\kappa \lambda} \delta^{\alpha}_{\mu}\delta^{\beta}_{\nu}  \no \\
& + 2 g^{\alpha \beta} \delta^{\kappa}_{\mu} \delta^{\lambda}_{\nu}   \Big], 
\end{eqnarray}
while the constants $a_{n}$, $b_{n}$, $c_{n}$, $p_{n}$, $q_{n}$, $r_{n}$, $s_{n}$ respectively
\begin{eqnarray}
& a_{n} \equiv \frac{2}{n-2} , \; \; \; \; b_{n} \equiv \frac{2}{(n-1)(n-2)}, \; \; \; \;    c_{n} \equiv \frac{1-2b_{n}}{1-a_{n}}, \\
& p_{n} \equiv a_{n} - \frac{a_{n}^2}{2b_{n}}(1+b_{n}) , \; \; \; \; q_{n} \equiv \frac{1}{4}[1 - a_{n}(a_{n} - 2b_{n})],    \\ 
& r_{n} \equiv 2[1 - \frac{a_{n}^2}{4}(n-4)]  ,  \; \; \; \;  s_{n} \equiv \frac{1}{2}[2a_{n}(a_{n} - 2b_{n})  - 1].
\end{eqnarray}

\end{appendix}

\end{document}